\documentclass[letter,longauth]{aa}  
\usepackage[utf8]{inputenc}
\usepackage{units}
\usepackage{amsmath}
\usepackage{url}
\usepackage{txfonts}
\usepackage{graphicx}
\usepackage{subfigure}
\usepackage{hyperref}
\usepackage{csquotes}
\usepackage{times}
\usepackage{natbib}
\usepackage{color}
\usepackage{ulem}
\usepackage[switch, modulo]{lineno}

\newcommand{\hess}{H.E.S.S.}

\newcommand{\RNum}[1]{\uppercase\expandafter{\romannumeral #1\relax}} 

\begin{document}

\title{\hess{}~Observations of the Crab during its March 2013 GeV Gamma-Ray Flare}

\author{H.E.S.S. Collaboration
\and A.~Abramowski \inst{1}
\and F.~Aharonian \inst{2,3,4}
\and F.~Ait Benkhali \inst{2}
\and A.G.~Akhperjanian \inst{5,4}
\and E.~Ang\"uner \inst{6}
\and G.~Anton \inst{7}
\and S.~Balenderan \inst{8}
\and A.~Balzer \inst{9,10}
\and A.~Barnacka \inst{11}
% ...
\email{kornelia.stycz@desy.de}}

\institute{
Universit\"at Hamburg, Institut f\"ur Experimentalphysik, Luruper Chaussee 149, D 22761 Hamburg, Germany \and
Max-Planck-Institut f\"ur Kernphysik, P.O. Box 103980, D 69029 Heidelberg, Germany \and
Dublin Institute for Advanced Studies, 31 Fitzwilliam Place, Dublin 2, Ireland \and
National Academy of Sciences of the Republic of Armenia, Yerevan  \and
Yerevan Physics Institute, 2 Alikhanian Brothers St., 375036 Yerevan, Armenia \and
Institut f\"ur Physik, Humboldt-Universit\"at zu Berlin, Newtonstr. 15, D 12489 Berlin, Germany \and
Universit\"at Erlangen-N\"urnberg, Physikalisches Institut, Erwin-Rommel-Str. 1, D 91058 Erlangen, Germany \and
University of Durham, Department of Physics, South Road, Durham DH1 3LE, U.K. \and
DESY, D-15738 Zeuthen, Germany \and
Institut f\"ur Physik und Astronomie, Universit\"at Potsdam,  Karl-Liebknecht-Strasse 24/25, D 14476 Potsdam, Germany \and
Nicolaus Copernicus Astronomical Center, ul. Bartycka 18, 00-716 Warsaw, Poland
% ...
}

\date{}

\abstract{On March 4, 2013 the \textit{Fermi}-LAT and AGILE reported a flare from the direction of the Crab Nebula
 in which the high-energy (HE; E~$>\unit[100]{MeV}$) flux was six times above its quiescent level. Simultaneous observations in other energy bands give us hints about the emission processes during the flare episode and the physics of pulsar wind nebulae in general.}
{We search for variability in the emission of the Crab Nebula at very-high energies (VHE; E~$>\unit[100]{GeV}$), using contemporaneous data taken with the \hess{}~array of Cherenkov telescopes. }
{Observational data taken with the \hess{}~instrument on five consecutive days during the flare were analysed for the flux and spectral shape of the emission from the Crab Nebula. Night-wise light curves are presented with energy thresholds of $\unit[1]{TeV}$ and $\unit[5]{TeV}$.}
{The observations conducted with \hess{}~on March 6 to March 10, 2013 show no significant changes in the flux. They limit the variation in the integral flux above $\unit[1]{TeV}$ to less than 63\% and the integral flux above $\unit[5]{TeV}$ to less than 78\% at a 95\% confidence level.}{}

\titlerunning{H.E.S.S. Observations of the March 2013 Crab Flare}
\maketitle
\section{Introduction}
The Crab Nebula (for an overview see \citealt{2008ARA&A..46..127H}) is a pulsar
wind nebula (PWN)
powered by
the Crab pulsar (in the following, the name \textit{Crab} is used synonymously for the system of the Crab pulsar and its nebula). 
The rotational energy of the pulsar is converted into kinetic energy of a relativistic pair-plasma flow terminating in a shock
with subsequent particle acceleration \citep{1974MNRAS.167....1R}. Unpulsed emission from the downstream flow (the nebula) covers all observable wavelengths.
The electrons and positrons of the plasma emit synchrotron radiation from radio
wavelengths up to several hundred MeV, and they Compton-upscatter ambient photons 
(see e.g.~\cite{1992ApJ...396..161D} and \cite{1996MNRAS.278..525A}) up to
energies of at least $\unit[80]{TeV}$ \citep{2004ApJ...614..897A}.

These processes manifest themselves as clearly distinguishable peaks in the spectral
energy distribution, which intersect in the energy band observed with
\textit{Fermi}-LAT \citep{2010ApJ...708.1254A}, AGILE \citep{2009A&A...502..995T} and EGRET \citep{2001A&A...378..918K}. Although the Crab 
is treated as a standard candle in very-high-energy (VHE; E~$>\unit[100]{GeV}$)
$\gamma$-ray astronomy (e.g.~\citealt{2010A&A...523A...2M}), its emission shows
substantial variability at high energies (HE; E~$>\unit[100]{MeV}$) (see
e.g.~\cite{2011Sci...331..736T, 2011Sci...331..739A, 2011ApJ...741L...5S,
2013ApJ...765...52S,0004-637X-749-1-26}), as well as at X-ray energies \citep{2011ApJ...727L..40W}, albeit with a smaller 
relative amplitude of flux changes ($\approx 5\%$) and on longer time scales of a few months.
 The most recent example is the flare
detected with \textit{Fermi}-LAT 
\citep{2013ATel.4855....1O, 2041-8205-775-2-L37} and AGILE \citep{2013ATel.4856....1S, 2013ATel.4867....1V} in March 2013, when the
peak photon flux of the synchrotron component above $\unit[100]{MeV}$ was (103.4~$\pm 0.8)\times
10^{-7}$~cm$^{-2}$~s$^{-1}$ compared to (6.1~$\pm 0.1)\times
10^{-7}$cm$^{-2}$~s$^{-1}$ in its quiescent state, and variability was measured on time scales of a few hours.

 As in previous flares (see e.g.~\cite{0004-637X-749-1-26}), the higher
flux state in March 2013 was accompanied by a hardening of the spectrum in the HE part of the synchrotron energy range. 
Generally, this implies either enhanced production
of electrons and positrons or changes in the magnetic and electric fields. While in the latter case, the 
inverse-Compton (IC) 
component will largely remain unchanged, in the former, the flare observed at a  synchrotron energy $E_\mathrm{syn}$ is accompanied by a flare
at a corresponding  energy E$_{IC}$ of IC scattered ambient photons. The apparent observed energy $E_\mathrm{syn}$ of
a few hundred MeV exceeds the maximum achievable energy of synchrotron radiation from shock-accelerated electrons/positrons
\citep{1983MNRAS.205..593G,1996ApJ...457..253D,2010MNRAS.405.1809L}. This observation indicates the presence of a mild Doppler boost or a different acceleration mechanism
altogether \citep{2010MNRAS.405.1809L,2013ApJ...770..147C}. Observations at the VHE band during flaring episodes 
 provide additional information on the conditions in the emission region (e.g. magnetic field, Doppler boost). In specific model scenarios, the relative variability expected at TeV
energies accompanying a major outburst at GeV energies ranges from $10^{-2}$ (see e.g. Fig.~8 in \cite{2011A&A...533A..10L}) to unity and higher (see e.g. 
\cite{2011MNRAS.414.2229B,2012MNRAS.424.2249K}). The detection of variability in the Crab Nebula with \hess{} is mainly limited by systematic uncertainties on the 
flux measurement of $\sim$ 20 -- 30$\%$. In addition, statistics rapidly decrease with increasing energy.

Given that the origin of the flares is poorly understood, the search for VHE counterparts of the flares is of great interest.
Moreover, the ARGO-YBJ group claimed nearly four times
higher event rates than average over a  period of eight days
\citep{2010ATel.2921....1A}  during a flare observed with AGILE \citep{2010ATel.2855....1T} and \textit{Fermi}-LAT
\citep{2010ATel.2861....1B} in September
2010. Whether the reported signals have an astrophysical origin that belongs to
the Crab Nebula remains unsettled, pending independent confirmation with other instruments. The contemporaneous observations of the Crab Nebula in March 2013 provide the opportunity to study the emission during a flaring state  at multiple wavelengths, ranging from infrared to X-rays \citep{2041-8205-775-2-L37} and VHE (\cite{2013arXiv1309.5949T} and H.E.S.S. observations reported in this paper). 
Spectral measurements at multi-TeV energies, which are most relevant in the search for an IC component, are required to complement our understanding of the flaring Crab Nebula and facilitate broadband modelling. The highest sensitivity for multi-TeV $\gamma$ rays is reached with ground-based Cherenkov telescope observations at high zenith angles, since the inclination angle of the induced air showers results in large effective areas. Since the Crab Nebula culminates at $\unit[45]{^\circ}$ for H.E.S.S., it provides the best observation conditions of all ground-based $\gamma$-ray telescopes.

\section{Data set and analysis}
The High Energy Stereoscopic System (\hess{}) is an array of five Imaging Air Cherenkov telescopes situated in the Khomas Highland, Namibia, at $\unit[1800]{m}$ above sea level. Since 2004, four telescopes (\hess{} Phase \RNum{1}) with mirror surfaces of $\sim\unit[100]{m^2}$ each have been detecting air showers produced by $\gamma$ rays with energies higher than $\unit[100]{GeV}$~\citep{2004NewAR..48..331H}. This array forms a square of $\unit[120]{m}$ side length. It has a field of view of $\unit[5]{^{\circ}}$ in diameter and a relative energy resolution of $\sim 14\%$ at $\unit[1]{TeV}$ \citep{2006A&A...457..899A}. In September 2012, a fifth telescope placed in the middle of the original square was inaugurated, initiating \hess{} Phase \RNum{2}. It has a mirror surface of $\sim\unit[600]{m^2}$ and lowers the energy threshold of \hess{} to tens of GeV.

Due to the flare, Fermi-LAT was switched to pointed target-of-opportunity observation mode of the Crab between MJD 56355 and 56359 \citep{2041-8205-775-2-L37}. The data presented here are ten observation runs taken in or shortly after this period, when the flux measured by Fermi-LAT was still about twice its average value. The data are comprised of runs with either three or four of the \hess{} \RNum{1} telescopes, each lasting 28 minutes. Since it was the rainy season in Namibia, observations were possible only during a few nights. In this period of time, the Crab Nebula was visible at large zenith angles for \hess{} (see
Table~\ref{table:runInfo}).

 The data were analysed with the \hess{}~Analysis
Package\footnote{HAP version hap-12-03-pl02} for shower reconstruction and a
multivariate analysis \citep{2009APh....31..383O} applying \textit{$\zeta$
std-cuts} for suppression of the hadronic background. To estimate the
cosmic-ray background, the \textit{reflected region} method
\citep{2007A&A...466.1219B} was used. Significances (in standard deviations,
$\sigma$) were calculated using Equation (17) in \citet{1983ApJ...272..317L}.
The analysis results for each night and for the whole data set can be found in
Table~\ref{table:runInfo}.  
A cross-check with an independent analysis \citep{2009APh....32..231D} and an independent data calibration indicates that
the systematic error on the flux normalisation is 30\% for this data set,
which is taken into account in the calculation of flux upper limits shown
below.

\section{Results}

\begin{table*}[]
\caption{Analysis results and for each night and the complete data set. Modified Julian date (MJD) of the start of the observation, live-time ($T_{\mathrm{live}}$), mean zenith angle ($Z_{\mathrm{mean}}$), the number of ON and OFF source events, the excess and its significance. The normalisation at $\unit[1]{TeV}$ (I$_0$) is given in units of  ($10^{-11}$cm$^{-2}$s$^{-1}$TeV$^{-1}$) and integral fluxes above $\unit[1]{TeV}$ and above $\unit[5]{TeV}$ in units of $10^{-11}$cm$^{-2}$s$^{-1}$. The underlying spectral model was assumed to be a power law. The given errors are statistical ones The estimated systematic errors are 30\% for all fluxes and 0.1 for spectral indices.
}

\begin{tabular}{cccccccccccc}
 \hline\hline

Date & MJD & $T_{\mathrm{live}}$ & $Z_{\mathrm{mean}}$ & $N_{\mathrm{ON}}$ & $N_{\mathrm{OFF}}$ & Excess &  Sign. & I$_0$ ($\unit[1]{TeV}$) & Index  & Flux $>\unit[1]{TeV}$ & Flux $>\unit[5]{TeV}$\\
2013    & -56300   & (s)        & (deg.)     &          &           &        & $\sigma$ & & &  & \\

\hline

03-06 & 57.8 & 3181 & 54 & 202 & 498 & 175 & 20 & $3.5 \pm 0.5 $ & $2.6 \pm 0.1 $ & $1.89 \pm 0.19 $ & $0.11 \pm 0.03 $ \\
03-07 & 58.8 & 3152 & 52 & 223 & 455 & 198 & 23 & $4.2 \pm 0.4 $ & $2.8 \pm 0.1 $ & $2.37 \pm 0.21 $ & $0.08 \pm 0.03 $ \\
03-08 & 59.8 & 3155 & 53 & 184 & 460 & 159 & 19 & $3.5 \pm 0.5 $ & $2.6 \pm 0.1 $ & $2.24 \pm 0.21 $ & $0.18 \pm 0.04 $ \\
03-09 & 60.8 & 4827 & 55 & 199 & 557 & 169 & 19 & $3.3 \pm 0.5 $ & $2.7 \pm 0.1 $ & $1.76 \pm 0.18 $ & $0.12 \pm 0.03 $\\
03-13 & 64.8 & 1596 & 54 & 62 & 173 & 53 & 11 & $5.2 \pm 1.4 $ & $3.4 \pm 0.3 $ & $2.06 \pm 0.36 $ & $0.06 \pm 0.05 $\\
full set & - & 15911 & 54 & 870 & 2143 & 754 & 42 & $3.8 \pm 0.2 $ & $2.7 \pm 0.1 $ & $2.14 \pm 0.10 $ & $0.12 \pm 0.01 $\\

\end{tabular}
\label{table:runInfo}
\end{table*}

Analysing the complete sample of ten runs taken in the nights from March 6
to March 10, 2013 (MJD 56358 - MJD 56365), we obtained an acceptance-corrected live
time of 4.4 hours, yielding 754 excess events from the source region. A
simple power law and an exponential cut-off power law were considered to model the
energy distribution, motivated by previous publications
\citep{2006A&A...457..899A}. Low statistics for E~$>\unit[10]{TeV}$, however, made it impossible to
distinguish between an exponential cut-off and a simple power law model. This is not a characteristic of this specific data set: A sample of ten runs on the Crab Nebula from another period with similar telescope participation did not allow any discrimination between a power law model and a power law model with an exponential cut-off, either.
Therefore, the numerically more stable power law model was adopted for all spectra and fitted in the energy range $\left[ 0.681 - 46.46\right]$ TeV. The energy spectrum
of the complete sample is shown in Fig.~\ref{fig:spectra}, together with the
exponential cut-off power law spectrum taken from \cite{2006A&A...457..899A}
as a reference. Night-wise data were fitted with a power law model as well, and all results and their statistical errors are compiled in Table~\ref{table:runInfo}. 
The spectral analysis results of both night-wise and complete samples agree with \cite{2006A&A...457..899A}, where an exponential cut-off power law was the best-fitting spectral model with I$_0$($\unit[1]{TeV}$)~=~($3.76 \pm 0.07$)~$\cdot~10^{-11}$cm$^{-2}$s$^{-1}$TeV$^{-1}$, $\Gamma_{\gamma} = 2.39 \pm 0.03$, and E$_\mathrm{cutoff}$~=~($14.3 \pm 2.1$)~TeV. 

To test for the
compatibility of this spectrum with the spectrum of the flare data set 
presented here, a $\chi^2$-test was conducted. Under the optimistic 
assumption of cancelling systematics between both data sets, the spectrum from 
\cite{2006A&A...457..899A} served as the null hypothesis for testing the photon spectrum above 
$\unit[1]{TeV}$, $\unit[5]{TeV}$, and $\unit[10]{TeV}$, resulting in $\chi^2$/ndf values of 32.6/31, 15.7/14, and 5.0/7, respectively. These values indicate no significant difference in the spectra. Due to the low statistics in the last bin of the spectrum (four ON events, one OFF event) a likelihood profile was calculated as described in \cite{2005NIMPA.551..493R}. With this method, a deviation of the last spectrum point from the expected flux according to \cite{2006A&A...457..899A} is about 2.5 $\sigma$, including neither systematic uncertainties nor the statistic uncertainties on the spectrum from \cite{2006A&A...457..899A}.

\begin{figure}[h!]
\includegraphics[width=\columnwidth]{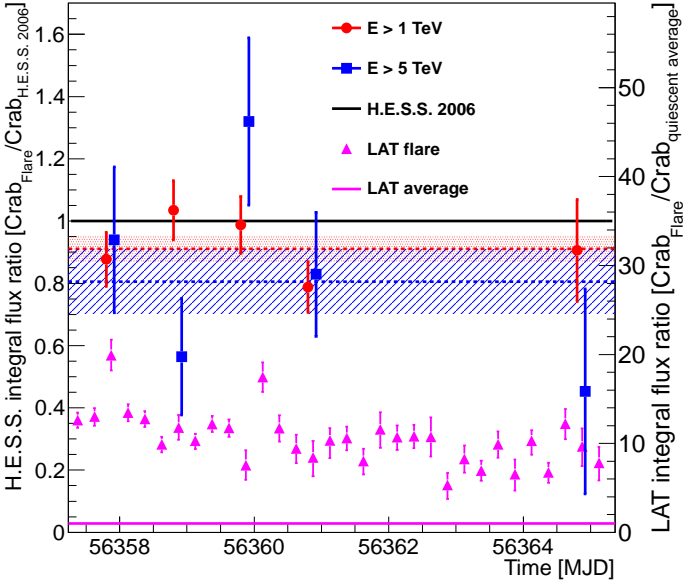}
\caption{Night-wise light curves for energy thresholds of 1 and $\unit[5]{TeV}$. Red 
squares indicate integral fluxes $>\unit[1]{TeV}$ relative to the integral flux above 
$\unit[1]{TeV}$ obtained from \cite{2006A&A...457..899A}. Error bars depict $\unit[1]
{\sigma}$ statistical errors. The dashed red line is the fit of a constant to this light 
curve, and the hatched red area marks the $\unit[1]{\sigma}$ statistical error. The equivalent 
data for an energy threshold of $\unit[5]{TeV}$ are presented in blue. For 
reference, the \textit{Fermi}-LAT synchrotron light curve as published in 
\cite{2041-8205-775-2-L37} is shown in magenta. Each bin corresponds to 6 hours of 
observations. The flux is scaled to the average quiescent synchrotron photon flux as 
reported in \cite{2012ApJ...749...26B} ((6.1 $\pm$ 0.2)$\cdot$ 10$^{-7}$ cm$^{-2}$ s$^{-1}$).}

\label{fig:NW_10TeV}
\end{figure}

\begin{figure}[h!]
\includegraphics[width=\columnwidth]{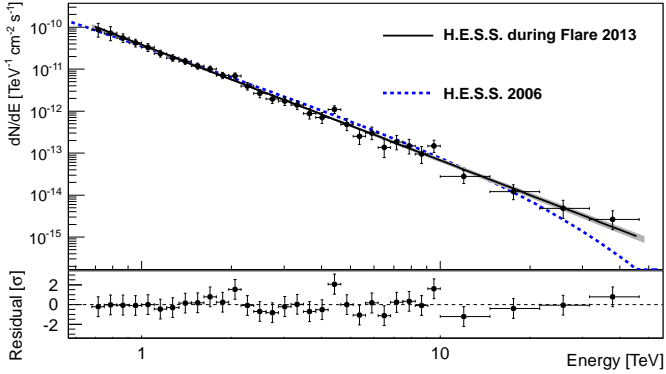}
\caption{Crab photon spectrum. Black circles indicate the \hess{}~Crab Nebula data taken in the nights from March 6 to March 10, 2013  with 1$\sigma$ error bars on the flux in the respective bin. The black line and the grey shaded area are the fitted power law model and  the corresponding 1$\sigma$ error butterfly. The blue dashed line corresponds to the spectrum reported in \cite{2006A&A...457..899A}.}
\label{fig:spectra}
\end{figure}

Since a flare in the MeV energy band is expected to be accompanied by an enhanced flux at tens of TeV \citep{2011A&A...533A..10L}, a search for variations in the flux above different energy thresholds was conducted. Integral fluxes above $\unit[1]{TeV}$ and  $\unit[5]{TeV}$ were calculated for the night-wise samples (see Fig.~\ref{fig:NW_10TeV}), and higher energy thresholds were tested but are non-restrictive owing to low statistics.
Fits of constants to the night-wise flux measurements give values of $(2.0\pm
0.1)\cdot~10^{-11}$cm$^{-2}$s$^{-1}$ with $\chi^{2}$/ndf = 6.1/4 and $(0.11 \pm
0.1)\cdot~10^{-11}$cm$^{-2}$s$^{-1}$ with $\chi^{2}$/ndf = 1.2/4 for an energy
threshold of $\unit[1]{TeV}$ and $\unit[5]{TeV}$, respectively. For comparison, the integral fluxes of the spectrum published in \cite{2006A&A...457..899A} above $\unit[1]{TeV}$ and above $\unit[5]{TeV}$ are $(2.26 \pm 0.08)\cdot~10^{-11}$cm$^{-2}$s$^{-1}$ and $(0.14 \pm 0.01)\cdot~10^{-11}$cm$^{-2}$s$^{-1}$, respectively.

The first night of \hess{}~observations (MJD 56358) is
coincident with the highest flux level detected by \textit{Fermi}-LAT in the March 2013 period of enhanced flux \citep{2041-8205-775-2-L37}. For that reason, upper limits on an enhancement of integral fluxes above $\unit[1]{TeV}$ and above $\unit[5]{TeV}$ were calculated for that night by comparison with the integral flux of the spectrum published in \cite{2006A&A...457..899A}. The spectrum in \cite{2006A&A...457..899A} was produced with a different analysis and under different observation conditions; therefore, event-number based upper limit calculations as put forward in \cite{2005NIMPA.551..493R} cannot be applied. Instead, the two flux values F$_{2006}$ and F$_{2013}$, determined by integration of the fitted spectral functions, are compared, which automatically takes energy migrations and efficiencies correctly into account. Since no significant deviation of F$_{2006}$ and F$_{2013}$ is found, and F$_{2006} > $  F$_{2013}$, a conservative 95\% confidence level upper limit is determined as F$_{2006} + 2\sigma$, where $\sigma$ comprises the quadratically added statistical and systematic errors.
With this method, the upper limit on an enhancement of the integrated flux above $\unit[1]{TeV}$ for the first night is 3.66~$\cdot~
10^{-11}$cm$^{-2}$s$^{-1}$ at a 95\% confidence level, corresponding
to an enhancement factor of 1.63 compared to the integrated flux published in 
\cite{2006A&A...457..899A}.
For the integrated flux above $\unit[5]{TeV}$, the upper limit on the flux enhancement factor relative to the integrated flux above $\unit[5]{TeV}$ as published in \cite{2006A&A...457..899A} is 1.78 at a 95\% confidence level. 

\section{Conclusions}
The upper limits on the enhancement of the Crab flux are far above what is expected for the TeV energy range from some models, which predict enhancement factors of at most 1.01, as described above referring to \cite{2011A&A...533A..10L}. In scenarios as in \cite{2011MNRAS.414.2229B} or \cite{2012MNRAS.424.2249K}, however, enhancement factors of 2 or more are possible, exceeding the upper limits presented here. Besides this, experimental evidence does exist for such a high relative flux variability: During the flare discovered by AGILE  in September 2010 \citep{2010ATel.2855....1T},
three to four times the average Crab flux at a mean energy of $\unit[1]{TeV}$ was reported by ARGO-YBJ for ten days with an observation time of about 5.5 hours  each \citep{2010ATel.2921....1A}. On July 3, 2012 ARGO-YBJ even observed an enhancement of eight times the average flux \citep{2012ATel.4258....1B} for a flare reported by Fermi on that day \citep{2012ATel.4239....1O}. Such an increase in flux  clearly lies above the upper limits presented in this paper and could be observed by the H.E.S.S. instrument if it was present during the observations at hand, rendering it unlikely.  More recently, the ARGO-YBJ group claimed a correlation of their Crab flux measurements with the varying \textit{Fermi}-LAT flux and an average flux enhancement factor of 2.4 $\pm$ 0.8  during flares at GeV energies \citep{2013arXiv1307.7041V}. This value is compatible with the 2$\sigma$ upper limits presented here only at the lower bound of its 1$\sigma$ errors.

On the other hand, both the MAGIC and VERITAS imaging atmospheric Cherenkov telescopes did not detect any flaring activity at VHE either during previous flares or during the period investigated in this paper. These instruments use observation times in units of $\sim$30 minutes, very similar to \hess{}~ For the flare in September 2010, both MAGIC and VERITAS did not detect any flux enhancement in 58 min during one night and 120 min during four nights, respectively \citep{2010ATel.2967....1M, 2010ATel.2968....1O}. For the flaring period discussed here, an integral flux above $\unit[1]{TeV}$  of $(2.05 \pm 0.07)\cdot~10^{-11}$cm$^{-2}$s$^{-1}$ was reported by VERITAS for a period of ten days with 10.3 hours of observations in total, compared to an integral flux of $(2.10 \pm 0.06)\cdot~10^{-11}$cm$^{-2}$s$^{-1}$ for observations outside the flare time window \citep{2013arXiv1309.5949T}. Taking the 30\% systematic error on flux measurements with VERITAS into account \citep{2013arXiv1309.5949T}, these numbers are in perfect agreement with the upper limits presented here and they give a very similar constraint on a possible flux enhancement.

The \textit{Fermi}-LAT energy spectra of the flaring component extending to energies of a few hundred MeV
favour at least a modest Doppler boosting. High angular resolution
observations of moving features in the nebula, however, do not show direct
evidence for bulk flow with $v>\unit[0.5]{c}$. It has been suggested that
modest Doppler factors could be realised at the region close to the termination
shock and that the optically resolved knot $\unit[0.6]{''}$ displaced from the
pulsar could be responsible
for the $\gamma$-ray variability
\citep{2011MNRAS.414.2017K}. In this scenario, the Doppler boost would lead to
an apparent enhancement of the inverse-Compton component for the stationary observer.
Not observing a transient feature at optical or X-ray frequencies \cite{2013ApJ...765...56W} during the flare 
is consistent with this picture given that the extrapolation of the observed $\gamma$-ray spectrum to lower energies would render
the X-ray/optical counterpart invisible against the bright nebula emission. Furthermore, a rather high value of the minimum energy
of the radiating electrons would basically lead to no sizeable emission at lower energies.

Assuming that the specific flux of the flare follows a power law $f_\nu\propto \nu^{-\alpha}$, the ratio of inverse-Compton and synchrotron emission at fixed frequencies 
scales with $f_\nu^\mathrm{IC}/f_\nu^{Syn}\propto (\delta/B)^{1+\alpha}$ \citep{1997ApJS..109..103D,2002A&A...388L..25G}, with $\delta$ the relativistic Doppler factor and $B$ the average magnetic field in the emission region. 
Therefore, the \hess{}~constraint combined with the contemporaneously measured \textit{Fermi}-LAT (synchrotron) flux
limits $\delta\lesssim 100 (B/122~\mu\mathrm{G})$. 

Future multi-wavelength measurements, especially with instruments with larger collection areas for TeV $\gamma$ rays like the planned Cherenkov telescope array, will be able to constrain such models even further.

\begin{acknowledgements}
\textit{ 
The support of the Namibian authorities and of the University of Namibia in facilitating the construction and operation of \hess{}~is gratefully acknowledged, as is the support by the German Ministry for Education and Research (BMBF), the Max Planck Society, the German Research Foundation (DFG), the French Ministry for Research, the CNRS-IN2P3 and the Astroparticle Interdisciplinary Programme of the CNRS, the U.K. Science and Technology Facilities Council (STFC), the IPNP of the Charles University, the Czech Science Foundation, the Polish Ministry of Science and  Higher Education, the South African Department of Science and Technology and National Research Foundation, and by the University of Namibia. We appreciate the excellent work of the technical support staff in Berlin, Durham, Hamburg, Heidelberg, Palaiseau, Paris, Saclay, and in Namibia in the construction and operation of the equipment.}
\end{acknowledgements}

\bibliographystyle{aa}
\bibliography{main}

\end{document}